\documentstyle[12pt,twoside,fleqn,espcrc1,epsfig]{article}

\newcommand{\beq}{\begin{equation}}
\newcommand{\eeq}{\end{equation}}
\newcommand{\bea}{\begin{eqnarray}}
\newcommand{\eea}{\end{eqnarray}}
\newcommand{\bce}{\begin{center}}
\newcommand{\ece}{\end{center}}
\newcommand{\etal}{{\it et al.}}

\newcommand{\AmS}{{\protect\the\textfont2
  A\kern-.1667em\lower.5ex\hbox{M}\kern-.125emS}}

\title{New Insight into Antiproton Production and Reabsorption Using Proton-Nucleus Collisions at the AGS} 

\author{Saskia Mioduszewski for the E910 Collaboration
\thanks{supported by US-DOE under contracts with BNL (DE-AC02-98CH10886), 
Columbia (DE-FG02-86ER40281), ISU (DE-FG02-92ER4069), KSU (DE-FG02-89ER40531),
LBNL (DE-AC03-76F00098), LLNL (W-7405-ENG-48), ORNL (DE-AC05-96OR22464), 
and UT (DE-FG02-96ER40982) and NSF under contract with FSU (PHY-9523974).}
\\ 
\vspace{0.3cm}
Department of Physics, Building 510C, Brookhaven National Laboratory, Upton, NY 11973, USA}

\begin{document}

\maketitle

\begin{abstract}
Antiproton ($\bar{p}$) yields are presented for proton-nucleus collisions, with 
targets Be, Cu, and Au, at beam momenta of 12.3 and 17.5 GeV/c.
In addition to target size and beam momentum, the number of projectile collisions
$\nu$, as derived from the number of ``grey'' tracks (slow protons and deuterons),
is used to disentangle the $\bar{p}$ reabsorption from the production.
By quantifying the amount of reabsorption of the $\bar{p}$ 
within the nucleus as a function of $\nu$, 
the annihilation within the nucleus is estimated and compared to the
free annihilation cross section.
Preliminary results on antilambda ($\bar{\Lambda}$) 
production as a function of $\nu$
are also presented for comparison.

\end{abstract}

\section{Introduction}

Sub-threshold $\bar{p}$ production as well as an apparently reduced 
$p-\bar{p}$ annihilation cross section in the nucleus have been under debate 
since the discovery of the $\bar{p}$ and until 
recently~\cite{subth1,subth2,bev,kb,sb,ARC,URQMD,RS00}. 
The observation of enhanced antimatter production has been proposed
as a signature of the Quark Gluon Plasma~\cite{hs}.  
Due to the annihilation of antibaryons in baryon-rich nuclear
matter, it has also been proposed to use $\bar{p}$ yields as a measure of the baryon
density in heavy ion collisions~\cite{gg}.  These interesting prospects for using
antibaryons to help determine the properties of the hot, dense phase in a heavy ion
collision require a deeper understanding of both the production and reabsorption
of the $\bar{p}$ within the nucleus. 
Proton-nucleus collisions provide a cleaner environment for testing
$\bar{p}$ production and reabsorption within the nucleus than heavy ion collisions.  
In this paper, we present measurements of $\bar{p}$ production in $p+$A collisions 
at the AGS that may help address the questions of production and reabsorption
in the nucleus.

\section{Data Reduction}
The E910 apparatus has been described elsewhere~\cite{e910}.
The time-of-flight (TOF) wall, used to identify the $\bar{p}$, is located
approximately 8~m from the target and covers approximately 5$\times$2~m$^2$.
Using the measured times of flight to identify particles, the $\bar{p}$ band 
is well separated from the pions and kaons up to 3.5 GeV/c.  
Momentum dependent cuts on the number of 
standard deviations of the measured TOF from the expected
TOF of a proton are applied.
To reduce background in the identified $\bar{p}$ sample, we apply cuts on the
particle's ionization energy loss in the TPC and the measured photoelectrons
in the Cerenkov detector.
Quality cuts on the hits on the TOF include a cut on
the difference in horizontal position between a projected track and the center of 
the hit TOF slat and a cut on the energy deposited on the TOF slat.
Tracks are matched to the TOF wall with a 90$\pm$5\% efficiency.
We estimate and subtract a momentum-dependent background of approximately 5\%.
Feeddown from $\bar{\Lambda}$ in our $\bar{p}$ sample is estimated to be
less than 5\%.
The data have been acceptance corrected within our $y-p_T$ coverage,
and corrected for the efficiencies of the cuts mentioned above.
All results are shown within our $y-p_T$ coverage, $y=(1,2)$ and $p_T=(10,800)$~MeV/c.

\section{Measured Antiproton Yields}
\begin{figure}[!htb]
\begin{minipage}[t]{9.5cm}
\vspace{-0.6cm}
\epsfig{figure=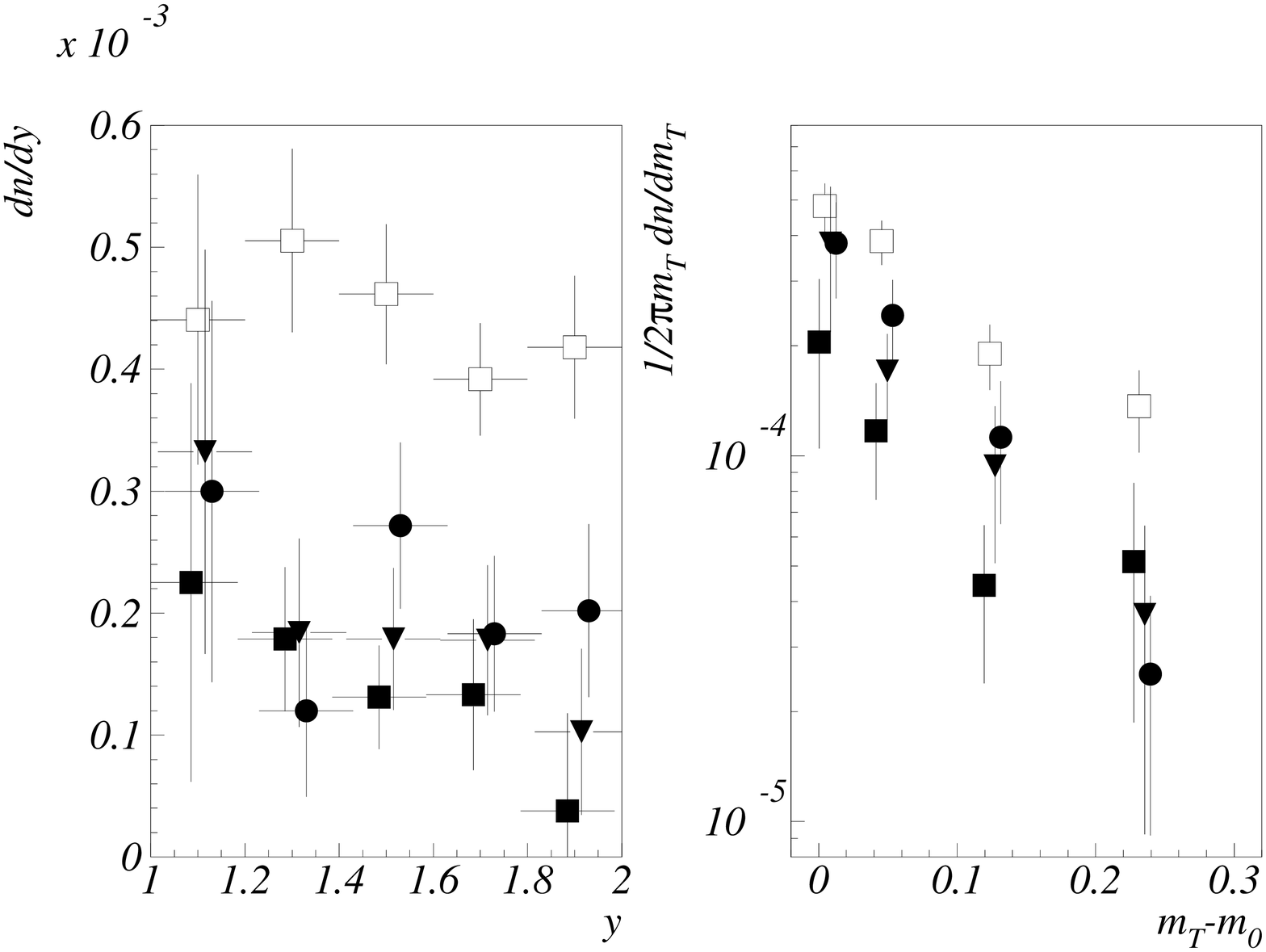,height=2.6in}
\vspace{-1.2cm}
\caption{Beam momentum and target dependence of rapidity densities (left) and 
transverse mass densities (right).  The 
open squares are 17.5~GeV/c $p+$Au yields, solid squares are 12.3~GeV/c $p+$Au,
solid triangles are 12.3~GeV/c $p+$Cu, and solid circles are 12.3~GeV/c $p+$Be.}  
\vspace{-0.3cm}
\label{data_target}
\end{minipage} \hfill
\hspace{0.4cm}
\begin{minipage}[t]{5.5cm}
\vspace{-0.3cm}
\epsfig{figure=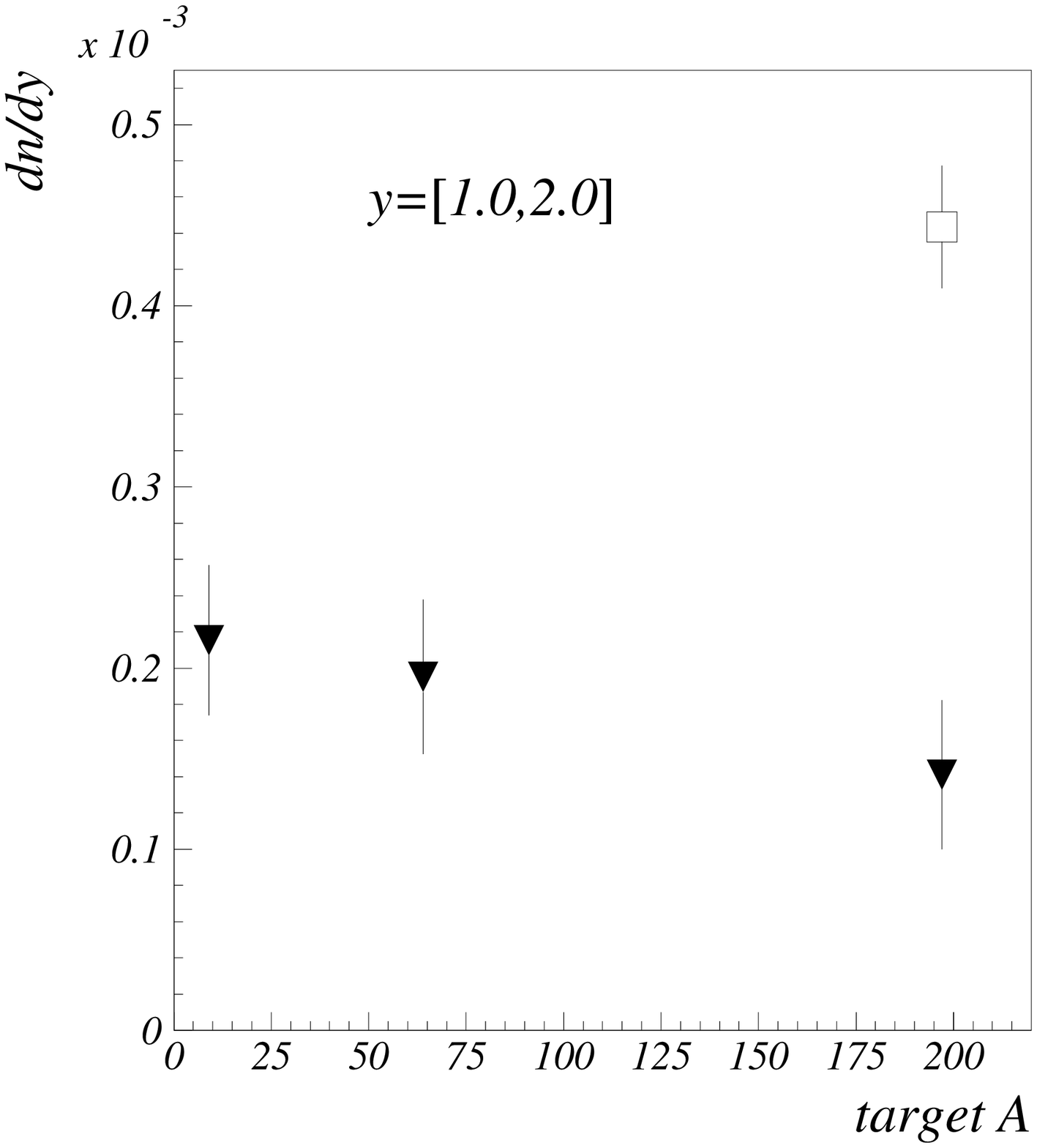,height=2.4in}
\vspace{-1.2cm}
\caption{Integrated rapidity density as a function of target A.  The triangles are 
yields from 12.3~GeV/c beam momentum, and the open square is 17.5~GeV/c.}  
\vspace{-0.3cm}
\label{data_int}
\end{minipage}
\end{figure}

The $\bar{p}$ yields are shown in Fig.~\ref{data_target}.  
We observe a strong increase in $p+$Au $\bar{p}$ yields from beam momentum 12.3 
to 17.5~GeV/c as expected, since production of $\bar{p}$ near threshold should depend
sensitively on the available phase space.
Although the likelihood of producing a $\bar{p}$ may be greater in a
larger nucleus~\cite{kd}, the likelihood of reabsorption is also greater
in the presence of more baryons.
These two countervailing effects can be studied
by investigating the target dependence of $\bar{p}$ yields.  Results 
for Be, Cu, and Au at beam momentum 12.3~GeV/c are also shown in
Fig.~\ref{data_target}.

Figure~\ref{data_int} shows the integrated rapidity densities for all four data sets.
The yields decrease from $p+$Be to $p+$Au collisions by 34~$\pm$~22\%.

\section{Reabsorption of the Antiprotons}
\begin{figure}[!htb]
\begin{minipage}[t]{7.5cm}
\vspace{-0.6cm}
\epsfig{figure=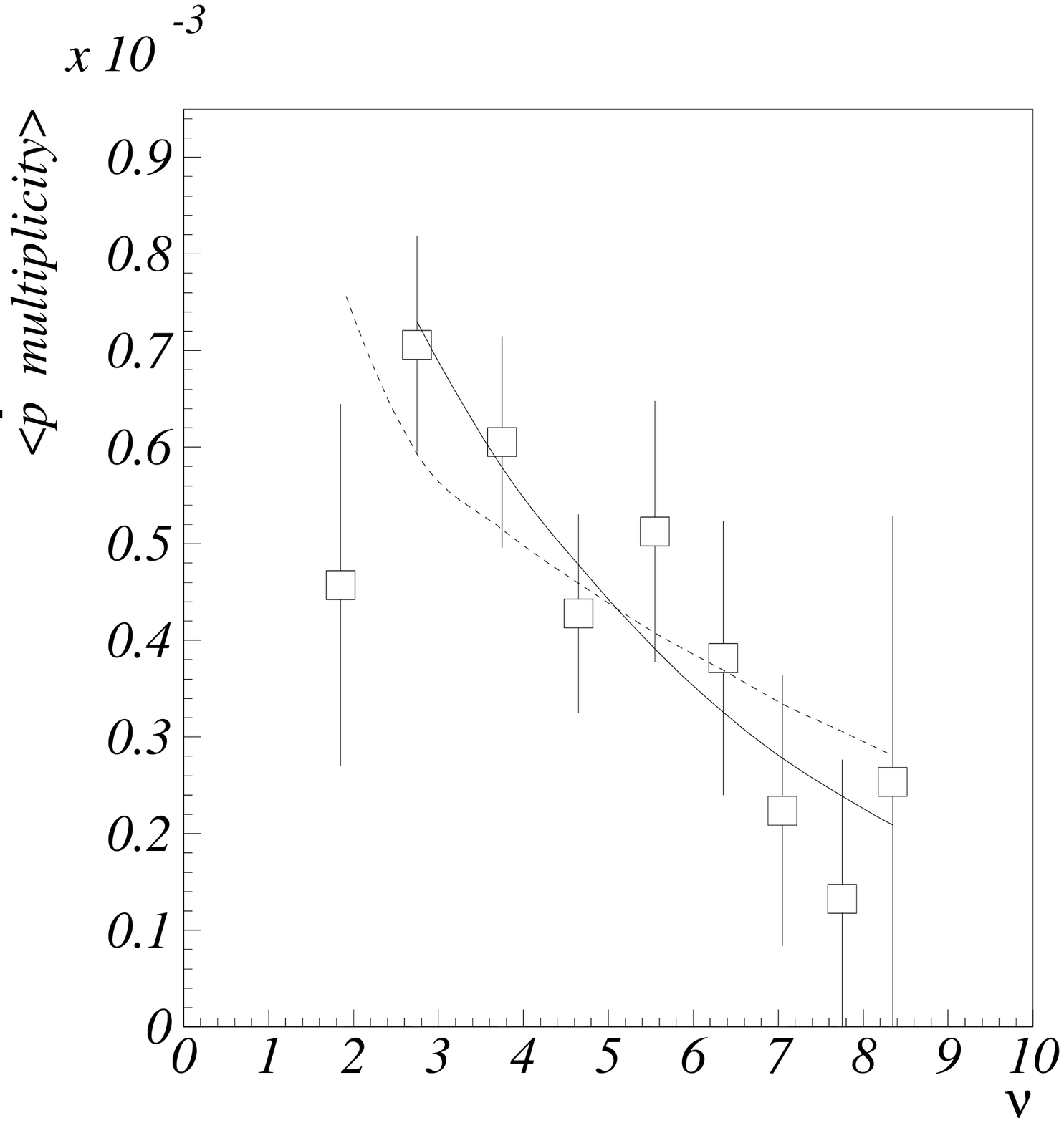,height=2.8in}
\vspace{-1.2cm}
\caption{Dependence of mean $\bar{p}$ yield on $\nu$.  The solid line is the result 
of the fit neglecting the first point ($N_g=0$ bin), and the dashed line includes 
the first point.}
\vspace{-0.3cm}
\label{au18_nu}
\end{minipage}\hfill
\begin{minipage}[t]{7.5cm}
\vspace{-0.6cm}
\epsfig{figure=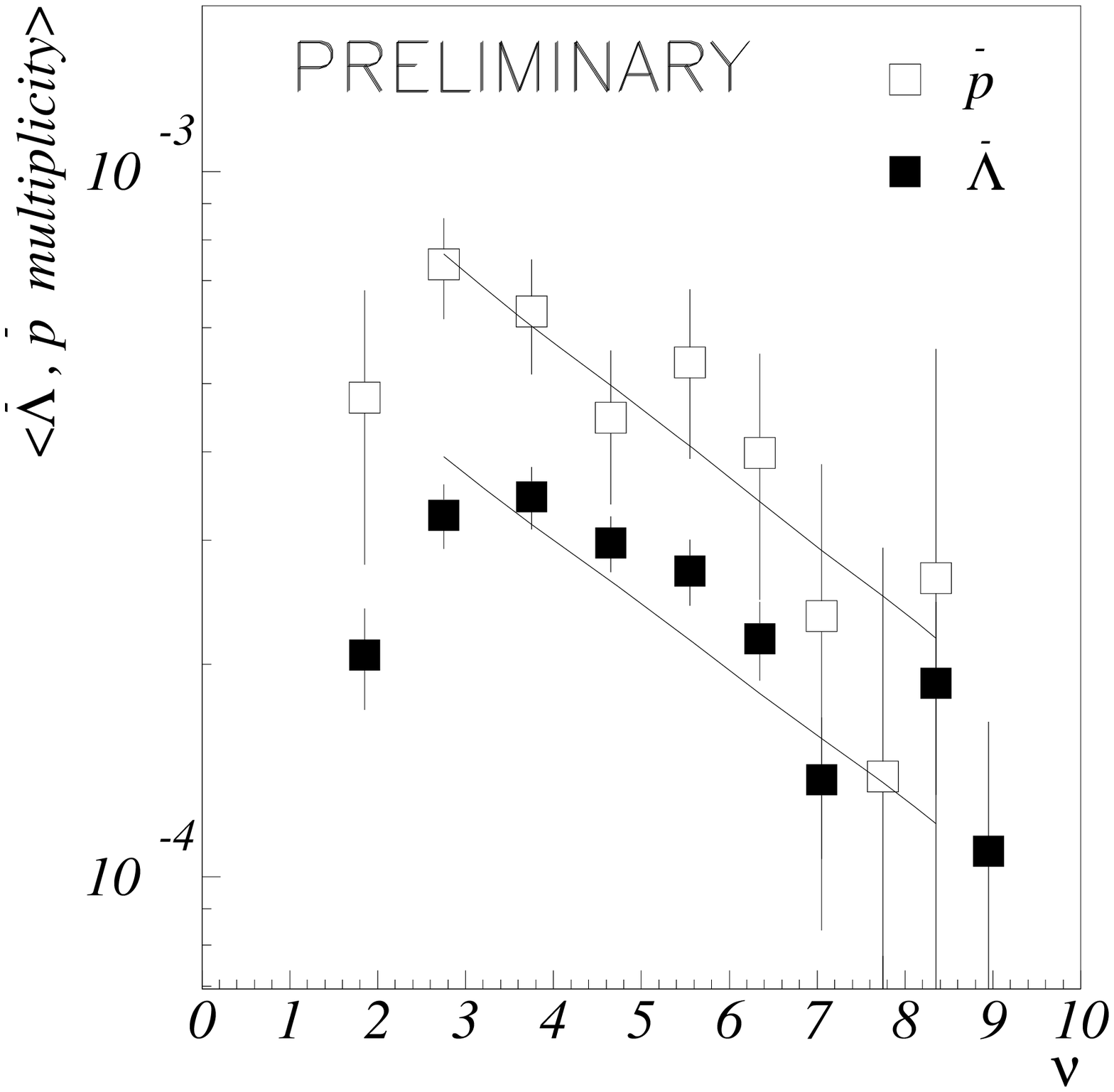,height=2.8in}
\vspace{-1.2cm}
\caption{Dependence of mean $\bar{p}$ yield and mean $\bar{\Lambda}$ yield on $\nu$ 
(semi-log scale).  
The open squares denote the $\bar{p}$ data and the solid squares denote the
$\bar{\Lambda}$ data.}  
\vspace{-0.3cm}
\label{au18_lambda_nu}
\end{minipage}
\end{figure}

By characterizing collision ``centrality,'' E910 can provide new insight into
$\bar{p}$ absorption.  
Events are characterized by the mean number of collisions $\nu$ that 
the projectile undergoes within the nucleus 
(as determined by the number of ``grey'' tracks $N_g$)~\cite{e910}.  
The $\nu$ dependence of the mean $\bar{p}$ multiplicity in 17.5~GeV/c $p+$Au 
collisions is shown in Fig.~\ref{au18_nu}.  A preliminary measurement of 
the mean $\bar{\Lambda}$ multiplicity as a function of $\nu$ is also shown in 
Fig.~\ref{au18_lambda_nu}.
The mean multiplicity of both tends to decrease as $\nu$ increases.
Although not convincingly significant, 
the increase from $N_g=0$ to $N_g=1$ in the mean $\bar{p}$ yield
may be evident of a contribution to production beyond the first $p+$N 
collision.  The increase is
more pronounced in the mean $\bar{\Lambda}$ yield versus $\nu$ and thus strengthens 
the evidence for production beyond a first collision model.
With the following assumptions, we quantify the ``effective'' absorption cross section 
in the nucleus and show that it is greatly reduced relative to the
free $p-\bar{p}$ annihilation cross section.  The first assumption
is that the $\bar{p}$ is predominantly produced in the first $p+$N
collision.
Since the beam energy is near the production threshold, this is generally 
assumed to be true at AGS energies~\cite{e802}.
If there are contributions to
production beyond $\nu=1$, as we have conjectured, 
they are not large enough to change our conclusion dramatically. 
The second assumption is that the $\bar{p}$ follows the path of the
projectile through the nuclear matter.  This is also a reasonable assumption 
because we observe strongly forward-peaked angular distributions for the 
$\bar{p}$.
Then the survival probability of the $\bar{p}$ can be described by
the following equation (although one should note that a formation time is not taken into account by this description),
\begin{equation}
\sigma(pA \rightarrow \bar{p}X) = \sigma(pp \rightarrow \bar{p}X) e^{-{\sigma_{abs} \over \sigma_{pN}}(\nu - 1)}.
\end{equation} 
Since the value $\nu$ plotted on the x-axis of
Figs.~\ref{au18_nu} and~\ref{au18_lambda_nu} 
is simply an average value, $\bar{\nu}(N_g)$, and each value of $N_g$ actually 
has a distribution of $\nu$ values associated with it, $P_{N_g}(\nu)$, we 
fold the above exponential with $P_{N_g}(\nu)$.
We determine $\sigma_{abs}$ by fitting with,
\begin{equation}
\sigma(pA \rightarrow \bar{p}X) = \sigma(pp \rightarrow \bar{p}X) P_{N_g}(\nu) e^{-{\sigma_{abs} \over \sigma_{pN}}(\nu
 - 1)}.
\end{equation} 
In one fit, the first data point is not included 
(because of the initial increase in yield from $N_g=0$ to $N_g=1$), 
and in the second fit, the $N_g=0$ point is included.
The parameter, $\sigma_{abs}/\sigma_{pN}$, 
resulting from the fit is $0.23~\pm~0.09$ when neglecting the first
data point in the fit, and $0.13~\pm~0.05$ when including it.  
Taking the more conservative estimate of 0.23 and assuming $\sigma_{pN}$ to
be 30~mb, one obtains an absorption cross section, $\sigma_{abs}$, of 
$6.9~\pm~2.7$~mb.
At $p=2.5$~GeV/c, the mean measured momentum 
of the $\bar{p}$ sample we detect, this is approximately $1/5$ of the free
annihilation cross section~\cite{kd}, $\sigma_{ann}$.  
The large discrepancy between $\sigma_{abs}$, as derived from our
model, and $\sigma_{ann}$ suggests a modification of the $p-\bar{p}$ 
annihilation cross section within the nuclear medium.
Figure~\ref{au18_lambda_nu} shows a very similar dependence of the mean
$\bar{\Lambda}$ yield on $\nu$.  Fitting with the same function that was used for the
$\bar{p}$ yields, the extracted fit parameter is $0.22~\pm~0.04$.  
The effective absorption cross section is thus the same (within errors) 
for $\bar{\Lambda}$ as for $\bar{p}$.
This suggests an intermediate state that emerges from the nuclear medium as
a $\bar{p}$ or a $\bar{\Lambda}$.

\section{Conclusions}
We have found that, at AGS energies, 
the $\bar{p}$ yields dramatically increase with beam momentum and moderately
decrease with increasing target size.
We have found evidence that even at these beam momenta, near the 
production threshold of the $\bar{p}$ and the $\bar{\Lambda}$, 
there is production beyond the first $p+$N collision for the
$\bar{\Lambda}$, and a similar behavior for the $\bar{p}$ is not excluded.  
Finally, the ``effective'' absorption cross section,
calculated within the context of a simple model, is significantly 
reduced relative to the free $p-\bar{p}$ annihilation cross section.
The similarity between the calculated absorption cross sections
for $\bar{p}$ and $\bar{\Lambda}$ may indicate the presence of a single
intermediate state which leads to both final states.

\vspace{0.3cm}


\end{document}